\documentclass{eptcs}  

\newif\ifeptcs
\eptcstrue		

\newif\ifieee
\ieeefalse		

\newif\ifllncs
\llncsfalse		

\newif\ifamsart
\amsartfalse             

\ifllncs\usepackage{smallsubsub}\fi
\ifieee
\usepackage{latex8}
\usepackage{times}
\fi

\usepackage{latexsym}
\usepackage{epic}
\usepackage{epsfig}
\usepackage{amsmath}
\usepackage{amssymb}
\usepackage{amscd}
\usepackage{url}
\usepackage{color}

\usepackage{graphics}

\input{xy}
\xyoption{all}

\newcommand{\cpsa}{\textsc{cpsa}}
\newcommand{\tagname}[1]{\ensuremath{\mathit{#1}\;}}
\newcommand{\kind}[1]{\ensuremath{\mathsf{#1}}}
\newcommand{\skind}[1]{\ensuremath{\mathtt{#1}}}

\newcommand{\Algebra}{\mathfrak{A}}

\newcommand{\dom}{\kind{dom}} \newcommand{\node}{\kind{nodes}}
\newcommand{\pk}[1]{\operatorname{\kind{pk}}(#1)}
\newcommand{\sk}[1]{\operatorname{\kind{sk}}(#1)}

\newcommand{\skel}{\ensuremath{\mathbb{A}}}
\newcommand{\skelB}{\ensuremath{\mathbb{B}}}

\ifamsart{}\else{

}\fi

\newcommand{\qdot}{\,\mathbf{.}\,}

\newcommand{\cons}{\,{\hat{\ }}\,}

\newcommand{\seq}[1]{\langle #1 \rangle}

\newcommand{\strandnode}[2]{ {#1}\downarrow{#2} }

\newcommand{\hull}{\kind{hull}}

\newcommand{\length}[1]{{| #1 |}}

\newcommand{\Hear}[2]{\kind{Lsn{#1}}[#2]}
\newcommand{\lang}{\ensuremath{\mathcal{L}}}

\newcommand{\enc}[2]{\{\!|#1|\!\}_{#2}}

\newcommand{\restricted}{\mathbin{\mid\!\hspace{-1.2pt}\grave{}}}

\newcommand{\restrict}[2]{#1 \restricted #2}

\newcommand{\subterm}{\sqsubseteq}
\newcommand{\term}{\kind{msg\/}}

\newcommand{\pleads}[2]{\ensuremath{{#2} (#1)}}

\newcommand{\uo}{\ensuremath{\mathsf{unique}}}
\newcommand{\no}{\ensuremath{\mathsf{non}}}

\newcommand{\applyrep}[2]{\ensuremath{#2 (#1)}}

\newcommand{\supfrown}{\ensuremath{{}^{\frown}}}

\ifllncs
\newtheorem{prop}{Proposition}{\bfseries}{\itshape}
\newtheorem{mylemma}[prop]{Lemma}{\bfseries}{\itshape}
\newtheorem{thm}{Theorem}{\bfseries}{\itshape}
\newtheorem{eg}{Example}{\bfseries}{\upshape}
\newtheorem{cor}[prop]{Corollary}{\bfseries}{\itshape}

\institute{The MITRE Corporation and \\ Worcester Polytechnic Institute%
}
\else
{\newtheorem{prop}{Proposition}[section]

\newtheorem{definition}[prop]{Definition}

\newtheorem{lemma}[prop]{Lemma}
\newtheorem{thm}[prop]{Theorem}

\ifamsart{}\else

\newenvironment{proof}{\noindent\emph{Proof. }}{{\vrule width2pt height8pt}\par\medskip}
\newcommand{\qed}{}
\fi}
\fi

\ifeptcs
\def\squareforqed{\hbox{\rlap{$\sqcap$}$\sqcup$}}
\def\qed{\ifmmode\squareforqed\else{\unskip\nobreak\hfil
\penalty50\hskip1em\null\nobreak\hfil\squareforqed
\parfillskip=0pt\finalhyphendemerits=0\endgraf}\fi}

\newenvironment{proof}{\rm {\noindent \bfseries Proof:\quad}}{\hspace*{\fill}\qed}
\fi

\ifeptcs %
\providecommand{\event}{EXPRESS}
\providecommand{\volume}{}
\providecommand{\anno}{2009}
\providecommand{\firstpage}{1}
\usepackage{breakurl}
\fi

\hypersetup{colorlinks=true} 

\title{Security Theorems via Model Theory%
\thanks{Supported by the MITRE-Sponsored Research program.  Author's
  address:  guttman@\{mitre.org,wpi.edu\}.}}

\author{ Joshua D.~Guttman %
\ifeptcs
  \institute{The MITRE Corporation}
  \institute{Worcester Polytechnic Institute}
\fi
}
\def\titlerunning{Sec.~Thms.~via Model Theory}
\def\authorrunning{J.D.~Guttman}

\begin{document} 

\maketitle


\nocite{DoghmiGuttmanThayer07,DoghmiGuttmanThayer07b}

\begin{abstract}
  A model-theoretic approach can establish security theorems, which
  are formulas expressing authentication and non-disclosure properties
  of protocols.  Security theorems have a special form, namely
  quantified implications $\forall\vec{x}\qdot
  (\phi\supset\exists\vec{y}\qdot\psi)$.

  Models (interpretations) for these formulas are \emph{skeletons},
  partially ordered structures consisting of a number of local
  protocol behaviors.  \emph{Realized} skeletons contain enough local
  sessions to explain all the behavior, when combined with some
  possible adversary behaviors.

  We show two results.  (1) If $\phi$ is the antecedent of a security
  goal, then there is a skeleton $\skel_{\phi}$ such that, for every
  skeleton $\skelB$, $\phi$ is satisfied in $\skelB$ iff there is a
  homomorphism from $\skel_{\phi}$ to $\skelB$.  (2) A protocol
  enforces $\forall\vec{x}\qdot (\phi\supset\exists\vec{y}\qdot\psi)$
  iff every realized homomorphic image of $\skel_{\phi}$ satisfies
  $\psi$.

  Since the program \textsc{cpsa} finds the minimal realized
  skeletons, or ``shapes,'' that are homomorphic images of
  $\skel_{\phi}$, if $\psi$ holds in each of these shapes, then the
  goal holds.
\end{abstract}


\section{Introduction}
\label{sec:intro}

Much work has been done in recent years on cryptographic protocol
analysis.  A central problem is, given a protocol, to determine
whether a formula, expressing a security goal about its behaviors in
the presence of an adversary, is true.  If the protocol achieves the
goal, one would like some explanation why. If it does not achieve the
goal, one would like a counterexample.
A security goal is a quantified implication:
\begin{equation}
  \label{eq:goal:form}
  \forall\vec{x}\qdot (\phi_0\supset\exists\vec{y}\qdot\psi).  
\end{equation}
The hypothesis $\phi_0$ is a conjunction of atomic formulas describing
regular behavior.  The conclusion $\psi$ is a disjunction of zero or
more such conjunctions, i.e.~$\psi$ is $\bigvee_{1\le i\le k}\phi_i$.
When the $\phi_i$ describe desired behaviors of other regular
participants, who are intended to be peers in protocol runs, then this
goal is an \emph{authentication} goal.  It says that each protocol run
contains at least one peer execution from $k$ different possibilities
among which the protocol may allow the participants to choose.

When $k=0$, $\psi$ is the empty disjunction \texttt{false}.  If
$\phi_0$ mentions an unwanted disclosure, (\ref{eq:goal:form}) says
the disclosure cannot occur.  Hence, security goals with $k=0$ express
\emph{secrecy} goals.%
\footnote{We will use $\phi,\phi_i$, etc., for conjunctions of 0 or
  more atomic formulas, and $\psi$ for disjunctions $\bigvee_{1\le
    i\le k} \phi_i$ where $0\le k$.}

Our models are \emph{skeletons}, partially ordered sets of
\emph{regular strands}, i.e.~local behaviors of regular participants.
A skeleton $\skel$ defines a set of executions, namely executions in
which images of these strands can be found.  We use
$\skel,\,\sigma\models\Phi$ in the classical sense, to mean that the
formula $\Phi$ is satisfied in the skeleton $\skel$, when the variable
assignment $\sigma$ determines how variables free in $\Phi$ are
interpreted.

A skeleton $\skel$ is an \emph{execution} if it is \emph{realized}.
This means that the message transmissions in $\skel$---when combined
with possible adversary behavior---suffice to explain every message
received in $\skel$.  A \emph{counterexample} to a goal $G$ is a
realized skeleton $\mathbb{C}$ such that, for some variable assignment
$\sigma$, $\mathbb{C},\sigma\models \phi$, and for every extension
$\sigma'$ of $\sigma$, $\mathbb{C},\sigma'\models \lnot\psi$.
$\mathbb{C}$ is a counterexample to $G$ only if $\mathbb{C}$ is
realized, even though $\models$ is also well-defined for non-realized
skeletons.

We focus on the homomorphisms among models.  A \emph{homomorphism} is
a structure-preserving map, which may embed one skeleton into a larger
one; may identify one strand with another strand that sends and
receives similar messages; and may fill in more information about the
parameters to the strands.  As usual, homomorphisms preserve
satisfaction for atomic formulas.  Suppose $H$ is a homomorphism
$H\colon\skel\mapsto\skelB$.  And suppose $\skel,\,\sigma\models\phi$,
i.e. the skeleton $\skel$ satisfies the atomic formula $\phi$ under an
assignment $\sigma$, which maps variables occurring in $\phi$ to
values that may appear in $\skel$.  Then
$\skelB,\,H\circ\sigma\models\phi$.

This holds for conjunctions of atomic formulas also.  Thus, a security
goal $\forall\vec{x}\qdot (\phi_0\supset\exists\vec{y}\qdot\psi)$,
concerns the homomorphic images of skeletons satisfying $\phi_0$.  If
any homomorphic image $\mathbb{C}$ is realized, then $\mathbb{C}$
should satisfy the disjunction $\psi$, i.e. $\mathbb{C}$ should
satisfy at least one of the disjuncts $\phi_i$ for $1\le i\le k$.

We already have a method for constructing homomorphisms from a
skeleton $\skel$ to realized skeletons~\cite{DoghmiGuttmanThayer07}.
The Cryptographic Protocol Shapes Analyzer \textsc{cpsa} is a program
that---given a protocol $\Pi$ and a skeleton of interest
$\skel$---generates all of the minimal, essentially different realized
skeletons that are homomorphic images of $\skel$.  We call these
minimal, essentially different skeletons \emph{shapes}, and there are
frequently very few of them.

\paragraph{Main Results.}  We show how a single run of the search for
shapes checks the truth of a security goal.  To determine whether
$\Pi$ achieves a goal $G=\forall\vec{x}\qdot
(\phi_0\supset\exists\vec{y}\qdot\psi)$, we find the shapes for the
single skeleton $\skel_{\phi_0}$.  Two technical results are needed to
justified this.
\begin{itemize}
  \item For any security hypothesis $\phi_0$, a single skeleton
  $\skel_{\phi_0}$ characterizes $\phi_0$.  I.e., for all $\skelB$:
  $$\exists \sigma\qdot\;\skelB,\,\sigma\models\phi_0
  \quad\mbox{iff}\quad \exists
  H\qdot\;H\colon\skel_{\phi_0}\mapsto\skelB.$$
  \item There exists a realized $\mathbb{C}$ that is a counterexample
  to $G$ iff there exists some shape
  $H\colon\skel_{\phi}\mapsto\skelB$ where $\skelB$ provides a
  counterexample.  
\end{itemize}
%
%
%
Our main results suggest a recipe for evaluating a goal
$G=\forall\vec{x}\qdot (\phi\supset\exists\vec{y}\qdot\psi)$ for a
protocol $\Pi$.
\begin{enumerate}
  \item Construct the skeleton $\skel_\phi$.
  \item Ask \textsc{cpsa} what shapes are accessible in $\Pi$,
  starting from $\skel_\phi$.
  \item As \textsc{cpsa} delivers shapes, check that each satisfies
  some disjunct $\phi_i$.\label{step:check:shape}
  \item If the answer is no, this shape is a counterexample to $G$.
  \item If \textsc{cpsa} terminates with no counterexample, then $G$
  is achieved.  
\end{enumerate}
Since the problem is undecidable~\cite{DurginEtAl04}, it is also
possible that \textsc{cpsa} will not terminate.
Step~\ref{step:check:shape} is easy, since each $\phi_i$ is a
conjunction of atomic formulas, and each shape is a finite (typically
small) structure.  
%

\paragraph{The Language of Goals.}  For each protocol, we define a
first order language $\lang(\Pi)$, in which
formulas~(\ref{eq:goal:form}) are security goals.  $\lang(\Pi)$
expresses authentication and secrecy goals~\cite{Lowe97} for $\Pi$,
including ``injective agreement'', as adapted to strand
spaces~\cite{Guttman01}.%
\footnote{ $\lang(\Pi)$ does not express observational
  indistinguishability properties, or ``strong
  secrecy''~\cite{Blanchet04}.}
It talks only about the roles.  One can say which roles executed, and
how far they executed in partial executions, and with what parameters.
Saying that different roles executed with the same values for certain
parameters is important.

However, $\lang(\Pi)$ is carefully designed to limit expressiveness.
$\lang(\Pi)$ says nothing about the forms of messages, and there are
no function symbols for encryption or pairing.  The protocol $\Pi$
determines the forms of messages, so to speak behind $\lang(\Pi)$'s
back in the semantics.  Thus, $\lang(\Pi)$ need only stipulate the
underlying parameters, when describing what has happened.

A benefit of this approach is that related protocols $\Pi,\Pi'$ may
have similar languages, or indeed identical languages, when
corresponding roles use the same parameters to compose messages of
different concrete forms.  This makes the languages $\lang(\Pi)$
suited to analyze protocol transformations (as in~\cite{Guttman09a})
to determine when security goals are preserved.  We used them (with
inessential differences) in~\cite{Guttman09}, where we gave a
syntactic criterion that ensures the safety of combining pairs of
protocols.  When $\Pi_1,\Pi_2$ meet the criterion, then any
goal~(\ref{eq:goal:form}) in $\lang(\Pi_1)$ that $\Pi_1$ meets is
still achieved by $\Pi_1\cup\Pi_2$.  Combining $\Pi_1$ with $\Pi_2$ to
form $\Pi_1\cup\Pi_2$ is a simple sort of transformation of $\Pi_1$.

\paragraph{Some Related Work.}  To document a protocol meeting a
security goal, one might like to provide a proof, e.g.~in Paulson's
style~\cite{Paulson98}, or in the Protocol Composition
Logic~\cite{DattaEtAl05}.  One might also view a counterexample as a
syntactic object much like a proof.  Symbolic constraint solving
techniques (starting
with~\cite{FioreAbadi01,MillenShmatikov01,RusinowitchTuruani01}) treat
them in this way, using rules, including unification, to construct
them.  The ``adversary-centered'' approach of
Selinger~\cite{Selinger01} also leads to a proof-like treatment of
protocol counterexamples, and to a model-theoretic view of achieving
goals.  To show that a goal is met, one exhibits a model in which
axioms are satisfied, but the adversary's knowledge does not include
any intended secrets.  These axioms describe the behavior of the
regular (non-compromised) participants.  The model is a set that is
invariant under disclosures effected by the protocol, but in which the
secrets do not appear.

We give another model-theoretic approach to achieving goals, but from
a ``protocol-centered'' point of view rather than an
``adversary-centered'' one.  In contrast to Selinger, who expresses
authentication properties of a protocol $\Pi$ by means of secrecy
properties of an expanded protocol $\Pi'$, we represent secrecy
properties as the special case of authentication properties where
$k=0$, as indicated above.

Chein and Mugnier also use homomorphisms to evaluate the truth of
implications, e.g.~\cite{CheinMugnier92,CheinMugnier04}, in the
context of conceptual graphs.  However, that context is quite
different, since their \emph{formulas} are graph-like objects, whereas
our \emph{interpretations} are graph-like structures.  Our framework
is tuned to the specific case of cryptographic protocols; for
instance, there is no analog of ``realized'' in their framework.  

\paragraph{Structure of this Paper.}  We start with some examples of
protocol goals in a simple protocol that does not achieve all of them,
and a corrected protocol that does (Section~\ref{sec:examples}).  In
Section~\ref{sec:language}, we define the first order classical
languages $\lang(\Pi)$ that express security goals for each protocol
$\Pi$.  Section~\ref{sec:skeletons} defines skeletons and
homomorphisms between them, and gives a semantics for $\lang(\Pi)$
using skeletons.  We show next in Section~\ref{sec:characteristic}
that each conjunction $\phi$ of atomic formulas has a characteristic
skeleton.  In Section~\ref{sec:goals}, we show how to use the
characteristic skeletons to check security goals.

\paragraph{Strand Spaces.} A \emph{strand} is a (linearly ordered)
sequence of nodes $n_1\Rightarrow\ldots\Rightarrow n_j$, each of which
transmits or receives some message $\term(n_i)$.  A strand may
represent the behavior of a principal in a single local session of a
protocol, in which case it is a \emph{regular} strand of that
protocol, or it may represent a basic adversary activity.  Basic
adversary activities include receiving a plaintext and a key and
transmitting the result of the encryption, and receiving a ciphertext
and its matching decryption key, and transmitting the resulting
plaintext.

A \emph{protocol} $\Pi$ is a finite set of strands, which are the
\emph{roles} of the protocol.  A strand $s$ is an \emph{instance} of a
role $\rho\in\Pi$, if $s=\applyrep{\rho}\alpha$, i.e.~if $s$ results
from $\rho$ by applying a substitution $\alpha$ to parameters in
$\rho$.

Message $t_1$ is an \emph{ingredient} of $t_2$, written $t_1\subterm
t_2$, if $t_1$ is used to construct $t_2$ other than as an encryption
key; i.e.~$\subterm$ is the smallest reflexive, transitive relation
such that $t_1\subterm t_1 \cons t_2$, and $t_2\subterm t_1 \cons
t_2$, and $t_1\subterm \enc{t_1}{t_2}$.

A message $t$ \emph{originates} on a strand node $n$ if (1)
$t\subterm\term(n)$; (2) $n$ is a transmission node; and (3)
$m\Rightarrow^+n$ implies $t\not\subterm\term(m)$.  A value that
originates only once in an execution is \emph{uniquely originating},
i.e.~a freshly chosen value.  A value that originates nowhere in an
execution may nevertheless be used within regular strands to encrypt
or decrypt.  However, if the adversary uses a value to encrypt or
decrypt, then the key must have been received, and must therefore have
originated somewhere.  Hence \emph{non-originating} values represent
uncompromised long term keys.
For more detail, see the Appendix.


\section{Examples}
\label{sec:examples}

\begin{figure}
  \centering
  $$\xymatrix@C=24mm{
    A\ar[r]^{\enc{\enc{k}{\sk{A}}}{\pk{B}}}\ar@{=>}[d]
    & \qquad\null\qquad\ar[r]^{\enc{\enc{k}{\sk{A}}}{\pk{B}}} 
    & B\ar@{=>}[d] \\ 
    \bullet
    & \qquad\null\qquad\ar[l]_{\enc{s}k}
    & \bullet\ar[l]_{\enc{s}k}
  }$$  
  \caption{Blanchet's ``Simple Example Protocol''}
  \label{fig:blanchet:example}
\end{figure}
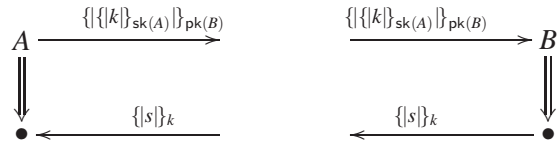

\paragraph{Blanchet's Example.}  We start from an example suggested by
Blanchet~\cite{Blanchet08}, as shown in
Fig.~\ref{fig:blanchet:example}.  A principal $A$ wishes to have $B$
transmit a secret to $A$ alone.
$A$ has a private signature key $\sk{A}$, with a public verification
key known to $B$.  $B$ has a private decryption key with a public
encryption key $\pk{B}$ known to $A$.  $A$ transmits a freshly chosen
symmetric key $k$ to $B$, signed by $A$ and encrypted for $B$.  $B$
then uses $k$ to encipher the secret $s$.

\paragraph{Authentication Goals of $A$.}  $A$ wants the protocol to
ensure that $s$ came from $B$.

To establish this, we attribute some assumptions to $A$, from which it
may follow that $B$ has transmitted $s$.  $A$ has had a run of her
side of the protocol, so the execution will contain the two nodes of
the strand shown on the left of the figure.  We must assume that $B$'s
private decryption key $\pk{B}^{-1}$ is uncompromised, in the sense
that this private key never originates.  It is thus used only by
regular participants in accordance with the protocol, and never by an
adversary.  Moreover, we assume that the symmetric key $k$ originates
only once, namely, on $A$'s first node.  In particular, the adversary
can not send it until after receiving $k$.  In a word, the adversary
will not \emph{guess} $k$.

$\skel_0$ summarizes these assumptions, as shown on the left in
Fig.~\ref{fig:blanchet:a:start}.  Here $\no_{\skel_0}$ is the set of
long term keys assumed uncompromised for the sake of this analysis,
and $\uo_{\skel_0}$ is the set of fresh values assumed uniquely
originating.  We call a diagram of this kind a \emph{skeleton}.
\begin{figure}
  \centering
    $$\color{blue}\framebox{\color{black}\xymatrix@C=8mm{
        A\ar[r]^{t_0}\ar@{=>}[d]
        & \null
        \\ 
        \bullet
        & \null\ar[l]_{\enc{s}k} \\
        \skel_0    
      }}
  \qquad\qquad \mapsto \qquad\qquad
  \framebox{\color{black}\xymatrix{
      A\ar[r]^{t_0}\ar@{=>}[d] 
      & \quad\prec\quad\ar[r]^{t_0} 
      & B\ar@{=>}[d]
      \\
      \bullet & \quad\succ\quad\ar[l]_{\enc{s}k}
      & \bullet\ar[l]_{\enc{s}k}    \\
      \skel_1
    }}
$$
  $$\{\pk{B}^{-1}\}=\no_{\skel_0}=\no_{\skel_1}, \{k\}=\uo_{\skel_0}=\uo_{\skel_1}$$
  \caption{Skeletons $\skel_0,\skel_1$, input and output for
    \textsc{cpsa}, with $t_0={\enc{\enc{k}{\sk{A}}}{\pk{B}}}$}
  \label{fig:blanchet:a:start}
\end{figure}

We want now to ``analyze'' this assumption, by which we mean, to find
all shapes accessible from $\skel_0$.  The shapes give all the
minimal, essentially different executions (realized skeletons)
accessible from $\skel_0$.  \textsc{cpsa} reports a single shape,
shown on the right of Fig.~\ref{fig:blanchet:a:start} as $\skel_1$.
This is what $A$ desired, as $B$ sent $s$ on its transmission node
at lower right.
 
The antecedent of the implication expressing this goal concerns the
starting point; is it $\phi^A=$
$$
\skind{Init2(n_2,a,b,k,s)} \land \skind{Unq}(\skind{k}) \land
\skind{Non}(\skind{sk(a)}) \land \skind{Non}(\skind{inv(pk(a))}).
$$
The predicate $\skind{Init2(n_2,a,b,k,s)}$ says that $\skind{n_2}$ is
the second node of an initiator strand with the given parameters.  The
remaining conjuncts express the supplementary assumptions about
non-compromised long term keys and a fresh session key.  The fact that
the only shape we obtain is $\skel_1$, which has a responder's second
node with the desired parameters, validates:
$$  
\phi^A \supset \exists \skind{m}\;\qdot\; \skind{Resp2(m,a,b,k,s)}. 
$$
%
%
The conclusion of this implication describes some of the additional
structure contained in $\skel_1$.

\paragraph{Confidentiality of $s$.}  The analysis for $A$'s
confidentiality goal is similar.  We again use the same assumptions,
augmented by the pessimistic assumption that $s$ is compromised.  We
represent this using a trick:  We use a \emph{listener node}
$\bullet\stackrel{s}\leftarrow$ that ``hears'' the value $s$ shorn of
any cryptographic protection.  Here we must also assume that
$s\in\uo_{\skel_2}$, since otherwise possibly the adversary will
simply guess (re-originate) $s$.  Thus, we start the analysis with the
form shown in Fig.~\ref{fig:blanchet:a:start:c}.
\begin{figure}[b]
  \centering
    $$\color{blue}\framebox{\color{black}\xymatrix@C=8mm{
    A\ar[r]^{t_0}\ar@{=>}[d]
    & \null
    \\
    \bullet & \null\ar[l]_{\enc{s}k} & \bullet\stackrel{s}\leftarrow }}
  \qquad\qquad \not\mapsto \qquad\qquad 
  {\color{black} \cdot}
  $$
  $$\{\pk{B}^{-1}\}=\no_{\skel_0} \qquad \{k,s\}=\uo_{\skel_0}$$
  \caption{$\skel_2$, for $A$'s confidentiality analysis, with no
    shapes }
  \label{fig:blanchet:a:start:c}
\end{figure}
We now reach an impasse:  \textsc{cpsa} reports that no shape exists,
starting from $\skel_2$.  Thus, $\skel_2$ is \emph{dead}:  No realized
skeleton can result from it.  We express this confidentiality goal in
the form:
$$ \phi^A\land\skind{Unq}(\skind{s})\land\skind{Lsn(m,s)} \supset
\skind{false} .$$ 
The conjunct $\skind{Lsn(m,s)}$ describes the listener node
$\skind{m}$ that ``hears'' the value $\skind{s}$.  This listener node
is incompatible with the other assumptions $\phi^A$.  Thus, $A$
achieves her goals using this protocol.

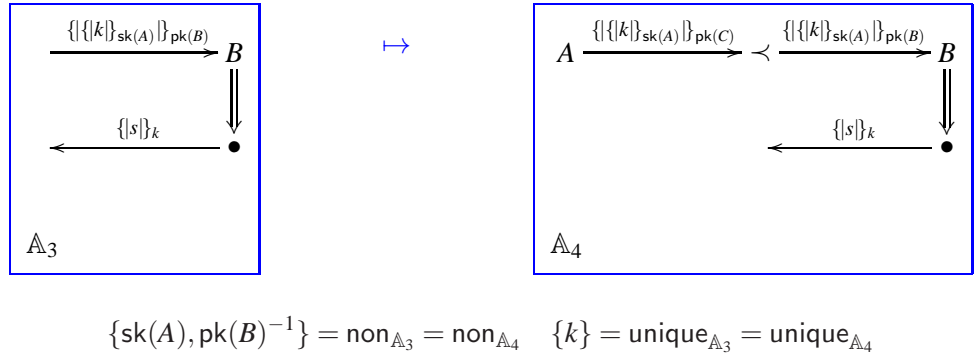
\begin{figure}
  \centering
    $$\color{blue}\framebox{\color{black}\xymatrix@C=20mm{
        \null\ar[r]^{\enc{\enc{k}{\sk{A}}}{\pk{B}}} 
        & B\ar@{=>}[d]
        \\ 
        \null  
        & \bullet\ar[l]_{\enc{s}k} \\
        \skel_3
  }}
  \qquad\qquad \mapsto \qquad\qquad 
  \framebox{\color{black}\xymatrix@C=20mm{
      A\ar[r]^{\enc{\enc{k}{\sk{A}}}{\pk{C}}}
      & \prec\ar[r]^{\enc{\enc{k}{\sk{A}}}{\pk{B}}} 
      & B\ar@{=>}[d]
      \\ 
      & \null  
      & \bullet\ar[l]_{\enc{s}k} \\
      \skel_4
    }}
  $$
  $$\{\sk{A},\pk{B}^{-1}\}=\no_{\skel_3}=\no_{\skel_4}\quad
  \{k\}=\uo_{\skel_3}=\uo_{\skel_4}$$ 
  \caption{$\skel_3,\skel_4$, for $B$'s authentication goal}
  \label{fig:blanchet:b:start}
\end{figure}

\paragraph{Authentication Goal for $B$.}  Unfortunately, the situation
is less favorable for $B$.  We start the analysis with the skeleton
$\skel_3$, shown on the left in Fig.~\ref{fig:blanchet:b:start}.  It
represents the hypothesis $\phi^B$:
$$
\skind{Resp2(m,a,b,k,s)} \land \skind{Unq}(\skind{k}) \land
\skind{Non}(\skind{sk(a)}) \land \skind{Non}(\skind{inv(pk(a))})
$$
We obtain the form $\skel_4$ shown on the right in
Fig.~\ref{fig:blanchet:b:start}.
Unfortunately, we have learnt nothing about the recipient $C$ for whom
$A$ intended this key.  Possibly $\pk{C}^{-1}$ is compromised.  Thus,
the adversary can decrypt ${\enc{\enc{k}{\sk{A}}}{\pk{C}}}$ and use
$B$'s public key to construct ${\enc{\enc{k}{\sk{A}}}{\pk{B}}}$.
Thus, skeleton $\skel_4$ is realized, but validates only: 
$$
\phi^B\supset \exists \skind{n,c}\;\qdot\; \skind{Init1(n,a,c,k)}.
$$
The expected initiator has got to step~1 of a run with some
$\skind{c}$, who---the protocol ensures---has originated the key $k$.
However, since possibly $C\not=B$, confidentiality for $k$ and $s$ may
fail.

The details of the \textsc{cpsa} run make clear how to fix the
protocol.  This requires us to replace $t_0$ with ${\enc{\enc{k\cons
      B}{\sk{A}}}{\pk{B}}}$, which includes $B$'s identity under $A$'s
signature.  In fact, Blanchet~\cite{Blanchet08} makes a more
complicated suggestion, using the component ${\enc{k\cons A\cons
    B}{\sk{A}}}$.  However, the \textsc{cpsa} analysis is very
precise, and indicates that only the responder's identity needs to be
included inside the signature.



\section{Language}
\label{sec:language}

We now consider the logical representation of our example
authentication and confidentiality goals.

\paragraph{Role Predicates.}  We need to be able to describe the
different kinds of nodes that are present in a skeleton, namely the
initiator's first and second nodes, and the responder's first and
second nodes, each of which has a number of parameters.  We must be
able to express these parameters, because we know (e.g.) that whether
the initiator's first node has $B$ or $C$ as its responder identity
parameter makes all the difference.  On the other hand, the form of
the messages is defined in the protocol, and is thus irrelevant to
describing what goals are achieved.  Thus, for Blanchet's simple
example protocol, we suggest four protocol-specific predicates:
$$
\begin{array}{r@{\qquad}l}
  \skind{Init1(m,a,b,k)}       & \skind{Resp1(m,a,b,k)} \\
  \skind{Init2(m,a,b,k,s)}     & \skind{Resp2(m,a,b,k,s)} 
\end{array}
$$
Suppose that we are given a variable assignment $\sigma$ that
associates values $\sigma(\skind{m}),\sigma(\skind{a})$, etc.~to the
variables $\skind{m},\skind{a}$, etc.  Then we read
$\skind{Init1(m,a,b,k)}$ as asserting that $\sigma(\skind{m})$ is a
node, which is the first step of an initiator strand, and the
initiator is $\sigma(\skind{a})$, its intended responder is
$\sigma(\skind{b})$, and it has created the key $\sigma(\skind{k})$
for this session.  We read $\skind{Init2(m,a,b,k,s)}$ as asserting
that $\sigma(\skind{m})$ is a node, which is the second step of an
initiator strand, and the initiator is $\sigma(\skind{a})$, its
intended responder is $\sigma(\skind{b})$, the session key is
$\sigma(\skind{k})$, and the intended secret is $\sigma(\skind{s})$.
Thus, if $\skind{Init2(m,a,b,k,s)}$, then we know that $\skind{m}$ is
preceded by a node $\skind{n}$ such that $\skind{Init1(n,a,b,k)}$ is
true under the same $\sigma$.  Analogous explanations hold for the
responder role.

These role predicates are similar to those used by Cervesato,
Durgin et al.~\cite{CervesatoEtAl00,DurginEtAl04a} in multiset
rewriting.

We also need a role predicate $\skind{Lsn(m,v)}$, which says, under an
assignment $\sigma$, that $\sigma(\skind{m})$ is a listener node that
receives that value $\sigma(\skind{v})$.  The language $\lang(\Pi)$,
where $\Pi$ is Blanchet's example protocol, contains these five role
predicates.

\paragraph{Shared Vocabulary.}  All languages $\lang(\Pi)$ also
contain some additional predicates.
$\skind{Preceq}(\skind{m},\skind{n})$ expresses the causal partial
ordering $\preceq$.  $\skind{Col}(\skind{m},\skind{n})$ says that
$\sigma(\skind{m})$ and $\sigma(\skind{n})$ are collinear, i.e.~they
lie on the same strand.  $\skind{Non}(\skind{v})$ and
$\skind{Unq}(\skind{v})$ express the assumptions that
$\sigma(\skind{v})\in\no$ and $\sigma(\skind{v})\in\uo$, resp.

In Section~\ref{sec:examples}, we also used the function symbols
$\skind{sk}, \skind{pk}$, and $\skind{inv}$ to talk about the long
term keys signature keys of principals, their long term public
encryption keys, and the inverses of those keys.

This is the full language $\lang(\Pi)$ where $\Pi$ is Blanchet's
example protocol.  As it happens, the language $\lang(\Pi')$, for the
corrected version of the protocol, is identical.  The roles are the
same, each with the same number of nodes, and with the same
parameters.  Thus, nothing in the language needs to change.

Apparently, for every goal $G\in\lang(\Pi)$, if $G$
is achieved in Blanchet's example protocol $\Pi$, then $G$ is also
achieved in its correction $\Pi'$.  Moreover, additional formulas are
achieved in $\Pi'$.  A satisfactory theory of protocol transformation
should give ways to prove (or disprove) intuitions like this one.

\paragraph{The Languages $\lang(\Pi)$.}  For each protocol $\Pi$,
$\lang(\Pi)$ is a language for talking about its executions.  We use
typewriter font $\skind{x},\skind{m}$, etc.~for syntactic items such
as variables or predicates within the language.

Suppose that $\Pi$ has $r$ protocol-specific roles
$\{\rho_1,\ldots,\rho_r\}$, where each role $\rho_i$ is of length
$\length{\rho_i}$, and the listener role.  We let $\{\skind{RP}_{0,1},
\dots, \skind{RP}_{r,\length{\rho_r}}\}$ be a collection of
$1+\sum_i\length{\rho_i}$ predicate symbols.  $\skind{RP}_{0,1}$ is
the listener role predicate, which we will write as
$\skind{Lsn(m,v)}$, indicating that node $\skind m$ receives the value
$\skind v$.  Each remaining predicate $\skind{RP}_{ij}$ takes
parameters $\skind{(m,v_1,\ldots, v_k)}$ where the $j^{\mathrm{th}}$
node on role $\rho_i$, and its predecessors, have involved $k$
parameters.

We write $\kind{fv}(\Phi),\kind{bv}(\Phi)$ for the free and bound
variables of any formula $\Phi$, defined in the usual way.  The empty
disjunction $\bigvee_{i\in\emptyset} \phi_i$ is identical with
$\skind{false}$; a one-element disjunction or conjunction is identical
with its single disjunct or conjunct.
\begin{definition}
\label{defn:lang:pi}
  \begin{enumerate}
    \item $\lang(\Pi)$ is the classical first order quantified
    language with vocabulary:
    \begin{description}
      \item[Variables] (unsorted) ranging over messages and nodes;
      \item[Function symbols] $\skind{sk}$, $\skind{pk}$,
      $\skind{inv}$;%
\footnote{We call terms built with these unary function symbols
  \emph{key terms}.} 
      \item[Predicate symbols] equality $\skind{u}=\skind{v}$,
      falsehood $\skind{false}$ (no arguments), and:
      \begin{itemize}
        \item $\skind{Non}(\skind{v})$, and $\skind{Unq}(\skind{v})$;
%
        \item $\skind{Col}(\skind{m},\skind{n})$ and
        $\skind{Preceq}(\skind{m},\skind{n})$;
        \item One role predicate $\skind{RP}_{ij}$ for each
        $\rho_i\in\Pi$ and $j$ with $1\le j\le\length{\rho_i}$.

        The predicate
        $\skind{RP}_{ij}(\skind{m},\skind{v_1},\ldots,\skind{v_k})$
        for the $j^{\mathrm{th}}$ node on $\rho_i$ has as arguments: a
        variable $\skind{m}$ for the node, and variables
        $\skind{v_{\ell}}$ for each of the $k$ parameters that have
        appeared in any of $\rho_i$'s first $j$ messages.
      \end{itemize}
    \end{description}
    \item A \emph{security claim} $\phi$ is a conjunction of atomic
    formulas of $\lang(\Pi)$ such that two conditions hold:
    \label{clause:lang:sec:claim}
    \begin{enumerate}
      \item Any two role predicate conjuncts have different variables
      as their first arguments $\skind{n},\skind{n'}$.
      \label{clause:lang:sec:claim:just:one:role}
      \item If a conjunct is not a role predicate, then each variable
      or key term that appears as an argument to it also appears as
      argument to some role predicate
      $\skind{RP}_{ij}(\skind{n},\skind{t_1},\ldots,\skind{t_i})$.
      \label{clause:lang:sec:claim:some:role}
    \end{enumerate}
    \item A sentence $\forall \vec{x}\qdot(\phi\supset
    \exists\vec{y}\qdot\psi)$ is a \emph{security goal} if (1) the
    $\vec{x}$s and $\vec{y}$s are disjoint; (2) $\phi$ is a security
    claim; and (3) $\psi$ is a disjunction $\bigvee_i \phi_i$ of
    conjunctions $\phi_i$ of atomic formulas.
\end{enumerate}
\end{definition}
The conditions in
Clauses~\ref{clause:lang:sec:claim:just:one:role}--\ref{clause:lang:sec:claim:some:role}
on security claims $\phi$ allow us to construct a single skeleton
$\skel_{\phi}$ to characterize $\phi$ (Thm.~\ref{thm:char:skel}).
Without Clause~\ref{clause:lang:sec:claim:just:one:role}, we would
have to be rather careful in our choice of message algebras and
protocols to ensure that there is a single ``most general'' role that
applies when two roles can have common instances.  The role predicate
$\skind{RP}_{ij}$ in Clause~\ref{clause:lang:sec:claim:some:role}
serves as an implicit sort declaration for the variables $\skind{x}$
appearing in it.  The sorts of parameters within the role $\rho_i$
determines the set of values that may lead to true instances of this
atomic formula.

We have already illustrated three security goals earlier, in
Section~\ref{sec:examples}; or, more precisely, we have shown three
formulas whose universal closures are security goals.  Another
relevant example is the ``missing'' confidentiality goal that a
responder would have wanted, but was actually not achieved by $\Pi$
but only by its correction $\Pi'$.  It is the universal closure of:
$$
\skind{Resp2(n_2,a,b,k,s)}  \land \skind{Unq}(\skind{k}) \land
  \skind{Non}(\skind{sk(a)}) \land \skind{Non}(\skind{inv(pk(a))})
  \land \skind{Lsn(m,s)} \supset \skind{false} .
$$

\paragraph{Axiomatizing Protocols.}  $\lang(\Pi)$ is specifically
intended to limit expressiveness, and there is no way to axiomatize
the behavior of protocols within it.  However, the slightly larger
language $\lang^+(\Pi)$ appears sufficient to axiomatize protocol
behaviors, and derive security goals.  It adds to $\lang(\Pi)$:
\begin{description}
  \item[Function symbols] $\skind{concat(v_1,v_2)}$ and
  $\skind{enc(v_1,v_2)}$, representing the concatenation of two
  messages $v_1,v_2$ and the encryption of a message $v_1$ using a
  second message $v_2$ as key; and \\
  $\skind{msgAt(n_1)}$ returning the message transmitted or received
  on the node $n_1$;
  \item[Predicate symbols] $\skind{Xmit(n_1)}$ and $\skind{Rcv(n_1)}$,
  true if $n_1$ is a transmission node or reception node, resp.
\end{description}
A few inductively defined notions such as ``message $t_0$ is found
only within the set of encryptions $S$ in message
$t_1$''~\cite[extended version, Def.~6]{DoghmiGuttmanThayer07} must be
introduced using these primitives.  The property of a skeleton being
realized can then be expressed as a closed sentence.  With these
notions, the reasoning encoded in \textsc{cpsa} could be carried out
axiomatically, at least in theory, within $\lang^+(\Pi)$.  The
important theorems would be the security goals, which lie within the
sublanguage $\lang(\Pi)$.


\section{Skeletons, Homomorphisms, and Satisfaction}
\label{sec:skeletons}

Before we define the satisfaction relation
$\skel,\,\sigma\models\phi$, we must define the skeletons that we have
already worked with in Section~\ref{sec:examples}.  We start by
summarizing our assumptions about the message algebra; more detail may
be found in Appendix~\ref{sec:strands}.

\paragraph{Message Algebra.}  Let $\Algebra_0$ be an algebra equipped
with some operators and a set of homomorphisms
$\eta\colon\Algebra_0\rightarrow\Algebra_0$.  We call members of
$\Algebra_0$ \emph{atoms}.

For the sake of definiteness, we will assume here that $\Algebra_0$ is
the disjoint union of infinite sets of \emph{nonces}, \emph{atomic
  keys}, \emph{names}, and \emph{texts}.  The operator $\kind{sk}(a)$
maps names to (atomic) signature keys, and $K^{-1}$ maps an asymmetric
atomic key to its inverse, and a symmetric atomic key to itself.
Homomorphisms $\eta$ are maps that respect sorts, and act
homomorphically on $\kind{sk}(a)$ and $K^{-1}$.

Let $X$ is an infinite set disjoint from $\Algebra_0$; its
members---called \emph{indeterminates}---act like unsorted variables.
$\Algebra$ is freely generated from $\Algebra_0\cup X$ by two
operations: encryption $\enc{t_0}{t_1}$ and tagged concatenation
$\tagname{tag} t_0\cons t_1$, where the tags $\tagname{tag}$ are drawn
from some set $\mathit{TAG}$.  For a distinguished tag
$\tagname{nil}\!\!$, we write $\tagname{nil}\; t_0\cons t_1$ as
$t_0\cons t_1$ with no tag.  In $\enc{t_0}{t_1}$, a non-atomic key
$t_1$ is a symmetric key.  Members of $\Algebra$ are called
\emph{messages}.

A homomorphism $\alpha=(\eta,\chi)\colon\Algebra\rightarrow\Algebra$
consists of a homomorphism $\eta$ on atoms and a function $\chi\colon
X\rightarrow\Algebra$.  It is defined for all $t\in\Algebra$ by the
conditions:
\begin{quote}
  \begin{tabular}{r@{$\;=\;$}l@{\quad}l@{\qquad\qquad}r@{$\;=\;$}l}
    $\applyrep{a}{\alpha}$ & $\eta(a)$, & if $a\in\Algebra_0$ &
    $\applyrep{\enc{t_0}{t_1}}{\alpha}$ &
    $\enc{\applyrep{t_0}{\alpha}}{\applyrep{t_1}{\alpha}}$    \\
    $\applyrep{x}{\alpha}$ & $\chi(x)$, & if $x \in X$ & 
    $\applyrep{\tagname{tag} t_0\cons t_1}{\alpha}$ & 
    $\tagname{tag} \applyrep{t_0}{\alpha}\cons
    \applyrep{t_1}{\alpha}$
\end{tabular}
\end{quote}
Thus, atoms serve as typed variables, replaceable only by other values
of the same sort, while indeterminates $x$ are untyped.
Indeterminates $x$ serve as blank slots, to be filled by any
$\chi(x)\in\Algebra$.  Indeterminates and atoms are jointly
\emph{parameters}.

This $\Algebra$ has the most general unifier property, which we will
rely on.  That is, suppose that for $v,w\in\Algebra$, there exist
$\alpha,\beta$ such that $\applyrep{v}{\alpha}=\applyrep{w}{\beta}$.
Then there are $\alpha_0,\beta_0$, such that
$\applyrep{v}{\alpha_0}=\applyrep{w}{\beta_0}$, and whenever
$\applyrep{v}{\alpha}=\applyrep{w}{\beta}$, then $\alpha$ and $\beta$
are of the forms $\gamma\circ\alpha_0$ and $\gamma\circ\beta_0$.

\paragraph{Skeletons.}  A skeleton is
a partially ordered set of nodes, together with assumptions $\no$
about uncompromised long term keys and $\uo$ about freshly chosen
values.  We write $\strandnode{s}i$ for the $i^{\mathrm{th}}$ node
along $s$, using 1-based indexing.

A \emph{skeleton} $\skel$ consists of (possibly partially executed)
role instances, i.e.~a finite set of nodes, ${\node}({\skel})$, with
two additional kinds of information:
\begin{enumerate}
  \item A partial ordering $\preceq_{\skel}$ on ${\node}({\skel})$;
  \item Sets $\uo_{\skel},\no_{\skel}$ of atomic values assumed
  uniquely originating and non-originating in $\skel$.
\end{enumerate}
${\node}({\skel})$ and $\preceq_{\skel}$ must respect the strand
order, i.e.~if $n_1\in{\node}({\skel})$ and $n_0\Rightarrow n_1$, then
$n_0\in{\node}({\skel})$ and $n_0\preceq_{\skel} n_1$.  If
$a\in\uo_{\skel}$, then $a$ must originate at most once in
${\node}({\skel})$.  If $a\in\no_{\skel}$, then $a$ must originate
nowhere in ${\node}({\skel})$, though $a$ or $a^{-1}$ may be the key
encrypting some ingredient of $n\in{\node}({\skel})$.  

$\skel$ is a \emph{preskeleton} if it meets the conditions except that
some values $a\in\uo_{\skel}$ may originate more than once in
${\node}({\skel})$.  If $\skel$ is a preskeleton, and it is possible
to extract a skeleton by identifying nodes and atoms, then there is a
canonical, most general way to do so~\cite[extended version,
Prop.~6]{DoghmiGuttmanThayer07}.  The canonical skeleton extracted
from $\skel$ is called the \emph{hull} of $\skel$.  We write
$\hull_{\skel}$ for the homomorphism (Def.~\ref{def:homomorphism})
that maps a preskeleton $\skel$ to its hull.

A skeleton $\skel$ is a \emph{skeleton for} a protocol $\Pi$ if all of
its strands are strands of $\Pi$.  

$\skel$ is \emph{realized} if it can occur without additional activity
of \emph{regular} participants; i.e., for every reception node $n$,
the adversary can construct $\term(n)$ via the Dolev-Yao adversary
actions,%
\footnote{The Dolev-Yao adversary actions are:  concatenating messages
  and separating the pieces of a concatenation; encrypting a given
  plaintext using a given key; and decrypting a given ciphertext using
  the matching decryption key.}
using as inputs:
\begin{enumerate}
  \item the messages $\term(m)$ where $m\prec_{\skel}n$ and $m$ is a
  transmission node;
  \item indeterminates $x$; and
  \item any atomic values $a$ such that
  $a\not\in(\no_{\skel}\cup\uo_{\skel})$, or such that
  $a\in\uo_{\skel}$ but $a$ originates nowhere in $\skel$.
\end{enumerate}
\begin{definition}
\label{def:homomorphism}
Let $\skel_0,\skel_1$ be preskeletons, $\alpha$ a homomorphism on
$\Algebra$, and
$\zeta\colon{\node}_{\skel_0}\rightarrow{\node}_{\skel_1}$.
$H=[\zeta,\alpha]$ is a \emph{(skeleton) homomorphism} if
 \begin{enumerate}
   \item[$1a.$] For all $n\in\skel_0$,
   $\term(\zeta(n))=\applyrep{\term(n)}{\alpha}$, with the same
   direction, either transmission or reception;
   \item[$1b.$] For all $s,i$, if $\strandnode{s}{i}\in\skel$, then
   there is an $s'$ s.t.~for all $j\le i$, \quad
   $\zeta(\strandnode{s}{j})=\strandnode{s'}{j}$;
   \item[$2.$] $n\preceq_{\skel_0}m$ implies
   $\zeta(n)\preceq_{\skel_1}\zeta(m)$;
   \item[$3.$]
   $\applyrep{\no_{\skel_0}}{\alpha}\subseteq\no_{\skel_1}$;
   \item[$4a.$]
   $\applyrep{\uo_{\skel_0}}{\alpha}\subseteq\uo_{\skel_1}$;
   \item[$4b.$] If $a\in\uo_{\skel_0}$ and $a$ originates at
   $n\in\node_{\skel_0}$, then $\applyrep{a}{\alpha}$ originates at
   $\zeta(n)\in\node_{\skel_1}$.
 \end{enumerate}
We write $H\colon\skel_0\mapsto\skel_1$ when $H$ is a homomorphism
from $\skel_0$ to $\skel_1$.
When $\applyrep{a}\alpha=\applyrep{a}\alpha'$ for every $a$ that is an
ingredient or is used for encryption in $\dom(\zeta)$, then
$[\zeta,\alpha]=[\zeta,\alpha']$; i.e., $[\zeta,\alpha]$ is the
equivalence class of pairs under this relation.
\end{definition}
The condition for $[\zeta,\alpha]=[\zeta,\alpha']$ implies that the
action of $\alpha$ on atoms not mentioned in the $\skel_0$ is
irrelevant.  We write $H(n)$ for $\zeta(n)$ or $H(a)$ for
$\applyrep{a}{\alpha}$, when $H=[\zeta,\alpha]$.  Evidently,
preskeletons and homomorphisms form a category, of which skeletons and
homomorphisms are subcategory.

In Section~\ref{sec:examples}, we have already given examples of
homomorphisms.  Each of the shapes we have considered is a homomorphic
image of its starting point.  Thus, for instance, we have
homomorphisms $\skel_0\mapsto\skel_1$ and $\skel_3\mapsto\skel_4$.
$\skel_2$ is dead in the sense that there is no realized $\skelB$ such
that $\skel_2\mapsto\skelB$.

Our first {\cpsa} run in Section~\ref{sec:examples} tells us that
every homomorphism from $\skel_0$ to a realized skeleton goes ``by way
of'' $\skel_1$~\cite{DoghmiGuttmanThayer07}.  That is, if $\skelB$ is
realized and $H\colon\skel_0\mapsto\skelB$, then $H=H_1\circ H_0$
where $H_0\colon\skel_0\mapsto\skel_1$.  Thus, any realized skeleton
accessible from $\skel_0$ has at least the structure contained in
$\skel_1$, homomorphisms being structure-preserving maps.

In Figs.~\ref{fig:blanchet:a:start} and~\ref{fig:blanchet:b:start},
the homomorphism simply adds nodes.  I.e.~$\zeta$ is an embedding and
$\alpha$ is the identity.  However, in other homomorphisms, $\zeta$
may be a bijection and $\alpha$ does the work, mapping distinct values
to the same result.  In other cases $\zeta$ is non-injective, mapping
two distinct strands in source to the same strand in the target.  For
instance, suppose that $\skel$ is a preskeleton but not a skeleton,
because some $a\in\uo_{\skel}$ originates on two strands.  If the map
$\hull_{\skel}$ is well defined, then $\hull_{\skel}$ must map both
strands on which $a$ originates to the same strand in the target
skeleton.  Hence, non-trivial $\hull_{\skel}$ maps are examples of
non-injective homomorphisms.

\paragraph{Semantics for $\lang(\Pi)$.}  The semantics for
$\lang(\Pi)$ are classical, with each structure a skeleton for the
protocol $\Pi$.  This requirement builds the permissible behaviors of
$\Pi$ directly into the semantics without requiring an explicit
axiomatization.  

An \emph{assignment} $\sigma$ for $\skel$ is a partial function from
variables of $\lang(\Pi)$ to $\Algebra\cup\node(\Pi)$.  By convention,
if $\sigma$ is undefined for any variable $\skind{x}$ in
$\kind{fv}(\Psi)$, then $\skel,\,\sigma\not\models\Psi$.  We write
$\sigma_1\oplus\sigma_2$ for the partial function that---for
$\sigma_1$ and $\sigma_2$ with disjoint domains---acts as either
$\sigma_i$ on the domain of that $\sigma_i$.
\begin{definition}
  Let $\skel$ be a skeleton for $\Pi$.
  Extend any assignment $\sigma$ to key terms of $\lang(\Pi)$ via the
  rules:
  $\sigma(\skind{sk}(\skind{t}))=\kind{sk}(\sigma(\skind{t}))$,
  $\sigma(\skind{inv(t)})=(\sigma(\skind{t}))^{-1}$.

%
  \emph{Satisfaction.}  $\skel,\sigma\models{\Phi}$ is defined
  via the standard Tarski inductive clauses for the classical first
  order logical constants, and the base clauses:
    \begin{tabbing}
      $\skel,\sigma\models \skind{DblArrw}(\skind{m},\skind{n})$
      \quad\= iff\quad\= 
      $\sigma(\skind{v})\in\uo_{\skel}$ and $\sigma(\skind{v})$
      originates at node $\sigma(\skind{m})$;\kill 

      $\skel,\sigma\models \skind{u=v}$  \> iff\quad 
      $\sigma(\skind{u})=\sigma(\skind{v})$;
      \\
      $\skel,\sigma\models \skind{Non}(\skind{v})$  \> iff\quad
      $\sigma(\skind{v})\in\no_{\skel}$;
      \\
      $\skel,\sigma\models \skind{Unq}(\skind{v})$ \> iff\quad 
      $\sigma(\skind{v})\in\uo_{\skel}$;
      \\
      $\skel,\sigma\models \skind{Col}(\skind{m},\skind{n})$  \> iff\quad
      $\sigma(\skind{m}),\sigma(\skind{n})\in{\node}(\skel)$, and
      either 
      $\sigma(\skind{m})\Rightarrow^*\sigma(\skind{n})$ or
      $\sigma(\skind{n})\Rightarrow^*\sigma(\skind{m})$;  
      \\
      $\skel,\sigma\models \skind{Preceq}(\skind{m},\skind{n})$ \> iff\quad
      $\sigma(\skind{m})\preceq_{\skel}\sigma(\skind{n})$;\\[2mm]
      and, for each role $\rho_i\in\Pi$ and index $j$ on $\rho_i$, the
      predicate
      $\skind{RP}_{ij}(\skind{m},\skind{v_1},\ldots,\skind{v_k})$ 
      obeys the clause \\[2mm] 
      $\skel,\sigma\models
      \skind{RP}_{ij}(\skind{m},\skind{v_1},\ldots,\skind{v_k})$ \>\> iff \=
      $\sigma(\skind{m})\in{\node}(\skel)$, and\\
      \>\>\> $\sigma(\skind{m})$ is an instance of the $j^{\mathrm{th}}$ node on
      role $\rho_i$, \\ 
      \>\>\> with the parameters
      $\sigma(\skind{v_1}),\ldots,\sigma(\skind{v_k})$. 
    \end{tabbing}
    We write $\skel\models{\Phi}$ when $\skel,\sigma\models{\Phi}$ for
    all $\sigma$, e.g.~when $\Phi$ is a sentence satisfied by $\skel$.
\end{definition}
%
%
In protocols where there are two different roles $\rho_i,\rho_h$ that
differ only after their first $j$ nodes---typically, because they
represent different choices at a branch point after the
$j^{\mathrm{th}}$ node~\cite{GuttmanEtAl05,Froeschle08}---the two
predicates $\skind{RP}_{ij}$ and $\skind{RP}_{hj}$ are equivalent, as
Def.~\ref{def:role:instance} makes precise.

\begin{lemma}\label{lemma:homomorphism:preserves}
  Suppose $\skind{\phi}$ is an atomic formula, and
  $H\colon\skel\mapsto\skelB$.  If
  $\skel,\,\sigma\models\skind{\phi}$, then $\skelB,\,H\circ\sigma
  \models \skind{\phi}$.
\end{lemma}


\section{Characteristic Skeletons} 
\label{sec:characteristic}

We write $\restrict{\sigma}{\kind{fv}(\Phi)}$ for the partial function
$\sigma$ restricted in domain to the free variables of $\Phi$.
\begin{definition} \label{def:char:skel}
  A pair $\skel,\sigma_*$ is \emph{characteristic} for a formula
  $\Phi$ iff $\skel,\,\sigma_*\models\Phi$ and, for all
  $\skelB,\sigma$,
  \begin{equation}
    \label{eq:ch:skel}
    \skelB,\sigma\models\Phi \quad\mbox{implies }\quad 
    \exists ! H\qdot\; H\colon\skel\mapsto\skelB \;\mbox{ and }\;
    \restrict{\sigma}{\kind{fv}(\Phi)} = H\circ\sigma_*.
  \end{equation}
  If there is such a $\sigma_*$, then $\skel$ is a
  \emph{characteristic skeleton} for $\Phi$.
\end{definition}
Being a homomorphic image of this $\skel$ characterizes satisfiability
of $\Phi$.  $\skel$ has minimal structure needed to make $\Phi$ true,
in the sense that $\Phi$ is satisfiable in any $\skel'$ just in case
$\skel'$ results from $\skel$ by a structure-preserving map (a
homomorphism).  From the form of the definition, $\skel,\sigma_*$ is
universal among interpretations satisfying $\Phi$, and such a
$\skel,\sigma_*$ will be unique to within isomorphism.

%


\paragraph{Constructing a Characteristic Skeleton.}  In order to
construct a characteristic skeleton $\kind{cs}(\phi)$ for a security
claim $\phi=\bigwedge_{1\le i\le\ell}\phi_i$, we treat the successive
atomic formulas $\phi_j$ in turn.  As we do so, we maintain two data
structures.  One is an assignment $\sigma$ which summarizes what
atomic value or node we have associated to each variable we have seen
so far.  Initially $\sigma$ is the empty function.  The other is the
characteristic skeleton $\kind{cs}(\bigwedge_{1\le i\le{j-1}}\phi_i)$
constructed from the part of the formula seen so far.  This is
initially the empty skeleton.
If $\phi$ is unsatisfiable, then instead of returning
$\kind{cs}(\phi),\sigma$ we must fail.

We assume that the conjuncts of $\phi$ have been reordered if
necessary so that atomic formulas containing role predicates precede
atomic formulas of the other forms.  For convenience, we also
eliminate equations by replacing the left hand side by the right hand
side throughout the remainder of the formula.

%
%
\begin{description}
  \item[Base Case.]  If $\ell=0$, so that $\phi=\bigwedge_{1\le i\le
    0}\phi_i=\skind{true}$, then let $\kind{cs}(\phi)$ be the empty
  skeleton, and let $\sigma_0$ be the empty (nowhere defined)
  substitution.
  \item[Recursive Step.]  Let $\phi=\bigwedge_{1\le i\le
    \ell+1}\phi_i$, and let $\skel_{\ell}=\kind{cs}(\bigwedge_{1\le
    i\le \ell}\phi_i)$ be a characteristic skeleton for all but the
  last conjunct, relative to $\sigma_{\ell}$.  We take cases on the
  form of the last conjunct, $\phi_{\ell+1}$:
  \begin{description}
    \item[$\skind{RP}_{ij}(\skind{m},\skind{t_1},\ldots,\skind{t_k})$:\quad]
    By clause~\ref{clause:lang:sec:claim:just:one:role} in
    Defn.~\ref{defn:lang:pi}, the variable $\skind{m}$ is not in the
    domain of $\sigma_\ell$.  If variables appearing in
    $\skind{t_1},\ldots,\skind{t_k}$, are not in the domain of
    $\sigma_\ell$, select atoms of appropriate sorts, not yet
    appearing in $\skel_\ell$, letting $\sigma'$ be the result of
    extending $\sigma_\ell$ with these choices.

    Let $n_1\Rightarrow\ldots\Rightarrow n_j$ be the first $j$ nodes
    of the role $\rho_i$, instantiated with the values
    $\sigma'(\skind{t_1}),\ldots,$ $\sigma'(\skind{t_k})$.  If any
    $n_\lambda$ (with $1\le\lambda\le j$) originates any value
    $a\in\no_{\sigma_\ell}$, then we must fail.

    Otherwise, let the preskeleton $\skel'$ be the result of adding
    $n_1\Rightarrow\ldots\Rightarrow n_j$ to $\skel_\ell$, and let
    $\skel_{\ell+1}$ be its hull.  If the hull is undefined, fail.
    Otherwise, define
    $\sigma_{\ell+1}=(\hull_{\skel'}\circ\sigma')\oplus(\skind{m}\mapsto
    n_j)$.  Return $\skel_{\ell+1},\sigma_{\ell+1}$.
    \item[$\skind{Non}(\skind{t})$:\quad] By
    clause~\ref{clause:lang:sec:claim:some:role} in
    Defn.~\ref{defn:lang:pi}, $\sigma_\ell(\skind{t})=v$ is well
    defined.  If the result of adding $v$ to $\no_{\skel_{\ell}}$ is a
    skeleton, then this skeleton is $\skel_{\ell+1}$.  Otherwise, we
    fail.  Let $\sigma_{\ell+1}=\sigma_\ell$.
    \item[$\skind{Unq}(\skind{t})$:\quad] By
    clause~\ref{clause:lang:sec:claim:some:role} in
    Defn.~\ref{defn:lang:pi}, $\sigma_\ell(\skind{t})=v$ is well
    defined.  If the result of adding $v$ to $\uo_{\skel_{\ell}}$ is a
    preskeleton $\skel'$ whose hull is a well-defined skeleton, then
    this skeleton is $\skel_{\ell+1}$.  Otherwise, we fail.  Let
    $\sigma_{\ell+1}=\hull_{\skel'}\circ\sigma_\ell$.
    \item[$\skind{Preceq}(\skind{m},\skind{n})$:\quad] By
    clause~\ref{clause:lang:sec:claim:some:role} in
    Defn.~\ref{defn:lang:pi}, $\sigma_\ell(\skind{m})$ and
    $\sigma_\ell(\skind{n})$ are well defined.  Let $\skel_{\ell+1}$
    be $\skel_{\ell}$ with the ordering enriched so that
    $\sigma_\ell(\skind{m})\preceq_{\skel_{\ell+1}}\sigma_\ell(\skind{n})$,
    failing if the latter introduces a cycle because in fact
    $\sigma_\ell(\skind{n})\preceq_{\skel_\ell}\sigma_\ell(\skind{m})$.
    Let $\sigma_{\ell+1}=\sigma_{\ell}$.
    \item[$\skind{Col}(\skind{m},\skind{n})$:\quad] By
    clause~\ref{clause:lang:sec:claim:some:role} in
    Defn.~\ref{defn:lang:pi}, $\sigma_\ell(\skind{m})=\strandnode{s}k$
    and $\sigma_\ell(\skind{n})=\strandnode{s'}{k'}$ are well defined.
    If one strand, e.g.~$s$, is at least as long as the other, we
    would like to map the successive nodes of $s'$ to nodes of $s$.
    However, their messages and directions in ${\skel_{\ell}}$ may not
    be the same.  If the directions (transmit vs.~receive) conflict,
    then we must fail.  Otherwise, if the successive messages are
    unequal, we may succeed by unifying them.

    Let $\beta$ be the most general unifier such that, for each $i$
    where both $\strandnode{s}i$ and $\strandnode{s'}i$ are defined,
    $$\applyrep{\term(\strandnode{s}i)}\beta =
    \applyrep{\term(\strandnode{s'}i)}\beta. $$
    Let $\skel'$ be the preskeleton resulting from applying $\beta$
    throughout $\skel$, failing if this is impossible because any
    value in $\no_{\skel'}$ would originate somewhere.  Let $\skel''$
    be the preskeleton resulting from omitting $\applyrep{s'}\beta$,
    and identifying its nodes with those of $\applyrep{s}\beta$,
    failing if this identification introduces any cycle into the
    ordering.  
    If $\skel''$ does not have a well defined hull, then fail.
    Otherwise, let that hull be $\skel_{\ell+1}$.  Let
    $\sigma_{\ell+1}=\hull_{\skel''}\circ\beta\circ\sigma_\ell$.
  \end{description}

\end{description}

\begin{thm}\label{thm:char:skel}
  If a security claim $\phi=\bigwedge_{1\le i\le \ell}\phi_i$ is
  unsatisfiable, the procedure above fails.  If $\phi$ is satisfiable,
  then the procedure returns a pair $\skel,\sigma$ that is
  characteristic for $\phi$.
\end{thm}
\begin{proof} We follow the inductive definition of $\kind{cs}(\phi)$.
\begin{description}
  \item[Base Case.]  Let $\ell=0$, and $\phi=\skind{true}$.  Then
  $\kind{cs}(\skind{true})$ is the empty skeleton $\skel_0$.  In fact,
  every $\skelB$ satisfies $\skind{true}$ via the empty substitution,
  and there exists exactly one homomorphism
  $H\colon\skel_0\mapsto\skelB$.
  \item[Recursive Step.]  Let $\phi=\bigwedge_{1\le i\le
    \ell+1}\phi_i$, and let $\phi^{-}=\bigwedge_{1\le i\le
    \ell}\phi_i$ be its predecessor, with all but the last conjunct of
  $\phi$.  Let $\skel_{\ell}=\kind{cs}(\phi^{-})$ be a characteristic
  skeleton for all but the last conjunct, relative to $\sigma_{\ell}$.
  In particular, $\phi^{-}$ is satisfiable.  If $\phi$ is
  unsatisfiable, then we need to check we will fail at this step.  If
  this step succeeds, then we need to show that
  $\skel_{\ell+1},\sigma_{\ell+1}$ are characteristic for
  $\kind{cs}(\phi)$.

  Suppose that, for any $\skelB,\tau$, we have
  $\skelB,\tau\models\phi$.  Then $\skelB$ also satisfies $\phi^{-}$,
  so by the induction hypothesis, there is a unique homomorphism
  $H_{\ell}=[\zeta_{\ell},\alpha_{\ell}]\colon\skel_{\ell}\mapsto\skelB$
  to within isomorphism.  Moreover,
  $\restrict{\tau}{\kind{fv}(\phi^{-})}=\alpha_{\ell}\circ\sigma_{\ell}$.
  We take cases on the form of the last conjunct, $\phi_{\ell+1}$.  In
  each case, $H_{\ell}$ can be adjusted to form a unique homomorphism
  $H_{\ell+1}\colon\skel_{\ell+1}\mapsto\skelB$, to within
  isomorphism.

  We use the same notation as in the corresponding cases of the
  definition of $\kind{cs}$.  

  \begin{description}
    \item[$\skind{RP}_{ij}(\skind{m},\skind{t_1},\ldots,\skind{t_k})$:]
    This is not jointly satisfiable with $\phi^{-}$ iff the new nodes
    $n_1\Rightarrow\ldots\Rightarrow n_j$ originate a value
    $a\in\no_{\skel_\ell}$, or if the hull is not defined.  In these
    cases, $\kind{cs}$ fails.

    $\skelB,\tau\models\phi$, and, since
    $\skind{RP}_{ij}(\skind{m},\skind{t_1},\ldots,\skind{t_k})$ is its
    last conjunct $\skelB,\tau\models
    \skind{RP}_{ij}(\skind{m},\skind{t_1},\ldots,\skind{t_k})$.

    Thus, $\tau(\skind{m})$ is an instance of $\strandnode{\rho_i}{j}$
    with parameters $\tau(\skind{t_1}),\ldots,\tau(\skind{t_k})$.
    Naming $\tau(\skind{m})=n_j'$, we have
    $n_1'\Rightarrow\ldots\Rightarrow n_j'$, and by the construction
    of $\kind{cs}(\phi)$, we have $n_1\Rightarrow\ldots\Rightarrow
    n_j$ in $\kind{cs}(\phi)$.  Thus, we can extend the node map
    $\zeta_{\ell}$ to $\zeta_{\ell+1}$ by mapping each $n_\lambda$ to
    $n_\lambda'$.  This is the only extension compatible with $\tau$.  

    Moreover, for each new variable $\skind{v}$ appearing in
    $\skind{t_1}, \ldots, \skind{t_k}$, we extend $\alpha_{\ell}$ by
    sending $\sigma_{\ell+1}(\skind{v})$ to $\tau(\skind{v})$.  By the
    construction of $\sigma_{\ell+1}(\skind{v})$, this is a new value,
    not equal to any value mentioned in $\skel_\ell$.  So
    $\sigma_{\ell+1}$ is a partial function.  Moreover, there is no
    other way to extend $\sigma_{\ell}$ compatible with $\tau$.  The
    resulting $\alpha_{\ell+1}$, together with $\zeta_{\ell+1}$, forms
    a homomorphism $\skel_{\ell_1}\mapsto\skelB$, and is uniquely
    determined.
    \item[$\skind{Non}(\skind{t})$:] If not jointly satisfiable
    with $\phi^{-}$, then $\sigma_{\ell+1}(\skind{t})$ originates in
    $\skel_{\ell}$, and $\kind{cs}$ fails.

    Since $\skelB,\tau\models\phi_{\ell+1}$,
    $\tau(\skind{t})\in\no_{\skelB}$, so $H_{\ell}$ is also a
    homomorphism from $\skel_{\ell+1}$ to $\skelB$.
 
    \item[$\skind{Unq}(\skind{t})$:] The universality of the
    $\kind{hull}_{\skel'}$ homomorphism among homomorphisms to
    realized skeletons ensures that $H_{\ell}$ factors through
    $\kind{hull}_{\skel'}$.
    \item[$\skind{Preceq}(\skind{m},\skind{n})$:] Since
    $\skelB,\tau\models\phi_{\ell+1}$,
    $\tau(\skind{m})\preceq_{\skelB}\tau(\skind{n})$, so $H_{\ell}$ is
    also a homomorphism from $\skel_{\ell+1}$ to $\skelB$.
    \item[$\skind{Col}(\skind{m},\skind{n})$:] In the non-failing
    case, $\zeta_{\ell}$ maps $\sigma_{\ell}(\skind{m})$ and
    $\sigma_{\ell}(\skind{n})$ to nodes on the same strand.  Thus,
    $H_{\ell}$ factors through the homomorphism from $\skel_{\ell}$ to
    $\skel_{\ell+1}$.
  \end{description}

\end{description}

\end{proof}

 
\section{Security Goals}
\label{sec:goals}

We turn now to our second result, which puts the pieces together.

\begin{thm}
  Suppose that $G=\forall \vec{x}\qdot(\phi \supset \exists\vec{y}
  \qdot \psi)$ is a security goal in $\lang(\Pi)$ where $\phi$ is
  satisfiable.  

  $\Pi$ achieves $G$ iff, whenever
  $H\colon\kind{cs}(\phi)\mapsto\skelB$ is a shape, there is a
  $\sigma$ such that $\skelB,\,\sigma\models\psi$.
\end{thm}

\begin{proof}
  1.  Suppose that $\Pi$ achieves $G$.  By Thm.~\ref{thm:char:skel},
  $\kind{cs}(\phi)$ is well-defined.  By
  Lemma~\ref{lemma:homomorphism:preserves},
  $H\colon\kind{cs}(\phi)\mapsto\skelB$ implies that $\skelB$
  satisfies $\phi$.  If $\skelB$ is a shape, it is realized.  Since
  $\Pi$ achieves $G$, $\skelB$ satisfies $\psi$.

  2.  Suppose that $\Pi$ does not achieves $G$, so that there is a
  realized $\mathbb{C}$ which satisfies $\lnot G$.  Let
  $\psi=\bigvee_{1\le{i}\le\ell}\phi_i$.  Using the Tarski
  satisfaction clauses and the disjointness of $\vec{x},\vec{y}$, we
  obtain a $\sigma_{\mathbb{C}}$ such that
  $\mathbb{C},\,\sigma_{\mathbb{C}} \models \phi\land
  \bigwedge_{1\le{i}\le\ell}\lnot\phi_i$.

  Since $\mathbb{C},\,\sigma_{\mathbb{C}}\models\phi$, there is a $J$
  such that $J\colon\kind{cs}(\phi)\mapsto\mathbb{C}$, and
  $\restrict{\sigma_{\mathbb{C}}}{\kind{fv}(\phi)}=J\circ\sigma_*$.
  Using Prop.~8 of the extended version
  of~\cite{DoghmiGuttmanThayer07}, $J=K\circ H$ where
  $H\colon\kind{cs}(\phi)\mapsto\skelB$ is a shape.

  If this $\skelB$ satisfies any $\phi_i$, then so would $\mathbb{C}$
  by Lemma~\ref{lemma:homomorphism:preserves}.
\end{proof}

\paragraph{Conclusion.}  We have explained a way to ensure that a
protocol achieves a security goal $G$.  We use the antecedent $\phi$
to choose a skeleton, namely $\kind{cs}(\phi)$.  We then obtain the
shapes accessible from $\kind{cs}(\phi)$, e.g.~by using \textsc{cpsa}.
If any shape does not satisfy any disjunct of the conclusion of $G$,
then we have a counterexample.   If no counterexample is found, then
$G$ is achieved.  

In future work, we will apply this method to protocol transformation.
It suggests a criterion to ensure that the result $\Pi_2$ of a
protocol transformation preserves all goals achieved by its source
protocol $\Pi_1$.


\paragraph{Acknowledgments.}  I am grateful to my colleagues, Leonard
Monk, John Ramsdell, and Javier Thayer, for many relevant discussions.
Marco Carbone gave valuable comments.  John Ramsdell is the author of
the \textsc{cpsa} implementation.

{
\bibliography{../../inputs/secureprotocols/secureprotocols.bib}
\ifeptcs
\bibliographystyle{eptcs}
\else 
\ifieee
\bibliographystyle{latex8}               
\else 
\bibliographystyle{plain}               
\fi\fi
}

\appendix
\section{Messages and Protocols}
\label{sec:strands} 


Messages are abstract syntax trees in the usual way:
\begin{enumerate}
\item Let $\ell$ and $r$ be the partial functions such that for
  $t=\enc{t_1}{t_2}$ or $t=\tagname{tag}{t_1}\cons{t_2}$,
  $\ell(t)=t_1$ and $r(t)=t_2$; and for $t\in\Algebra_0$, $\ell$ and
  $r$ are undefined.
  \item A \emph{path} $p$ is a sequence in $\{\ell,r\}^{*}$.  We
  regard $p$ as a partial function, where $\seq{}=\kind{Id}$ and
  $\kind{cons}(f,p)=p\circ f$.  When the rhs is defined, we have:  1.
  $\seq{}(t)=t$; 2.  $\kind{cons}({\ell},p)(t)=p(\ell(t))$; and 3.
  $\kind{cons}({r},p)(t)=p(r(t))$.
  \item $p$ \emph{traverses a key edge} in $t$ if $\pleads{t}{p_1}$ is
  an encryption, where $p=p_1\supfrown \seq{r}\supfrown p_2$.
  \item $t_0$ \emph{is an ingredient of} $t$, written $t_0\subterm t$,
  if $t_0=\pleads{t}{p}$ for some $p$ that does not traverse a key
  edge in $t$.%
%
%
\item $t_0$ \emph{appears in} $t$, written $t_0\ll t$, if
  $t_0=\pleads{t}{p}$ for some $p$.
\end{enumerate}
%
%
%
A message $t_0$ \emph{originates} at a node $n_1$ if (1) $n_1$ is a
transmission node; (2) $t_0\subterm\term(n_1)$; and (3) whenever
$n_0\Rightarrow^{+}n_1$, $t_0\not\subterm\term(n_0)$.


In the tree model of messages, to apply a homomorphism, we walk
through, copying the tree, but inserting $\applyrep{a}\alpha$ every
time an atom $a$ is encountered, and inserting $\applyrep{x}\alpha$
every time that an indeterminate $x$ is encountered.
%
%
%
%

\paragraph{Protocols.} A \emph{protocol} $\Pi$ is a finite set of
strands which includes $\Hear{}{k}$, representing the roles of the
protocol.
%

A principal executing a role such as the initiator's role in
Fig.~\ref{fig:blanchet:example} may be partway through its run; for
instance, it may have executed the first transmission node without
``yet'' having executed its second event, the reception node.
\begin{definition}
  \label{def:role:instance}
  Node $n$ is a \emph{role node} of $\Pi$ if $n$ lies on some
  $\rho\in\Pi$.

  Let $n_j$ be a role node of $\Pi$ of the form
  $n_1\Rightarrow\ldots\Rightarrow n_j\Rightarrow\ldots$.  Node $m_j$
  is an \emph{instance} of $n_j$ if, for some homomorphism $\alpha$,
  the strand of $m_j$, up to $m_j$, takes the form:
  $\applyrep{n_1}{\alpha}\Rightarrow\ldots\Rightarrow
  \applyrep{n_j}{\alpha}=m_j$.   
\end{definition}
That is, messages and their directions---transmission or
reception---must agree up to node $j$.  However, any remainders of the
two strands beyond node $j$ are unconstrained.  They need not be
compatible.  When a protocol allows a principals to decide between
different behaviors after step $j$, based on the message contents of
their run, then this definition represents
branching~\cite{Froeschle08,GuttmanEtAl05}.  At step $j$, one doesn't
yet know which branch will be taken.



\end{document}